# Histopathological Imaging Classification of Breast Tissue for Cancer Diagnosis Support Using Deep Learning Models

Tat-Bao-Thien Nguyen[1], Minh-Vuong Ngo[2], Van-Phong Nguyen[3]

[1] Vietnam Aviation Academy, Ho Chi Minh City, Vietnam
thienntb@vaa.edu.vn
[2] Ho Chi Minh City Open University, Ho Chi Minh City, Vietnam
vuong.nm@ou.edu.vn
[3] University of Information Technology, VNU-HCM, Vietnam
phongnv.13@grad.uit.edu.vn

**Abstract.** According to some medical imaging techniques, breast histopathology images called Hematoxylin and Eosin are considered as the gold standard for cancer diagnoses. Based on the idea of dividing the pathologic image (WSI) into multiple patches, we used the window [512,512] sliding from left to right and sliding from top to bottom, each sliding step overlapping by 50% to augmented data on a dataset of 400 images which were gathered from the ICIAR 2018 Grand Challenge. Then use the EffficientNet model to classify and identify the histopathological images of breast cancer into 4 types: Normal, Benign, Carcinoma, Invasive Carcinoma. The EffficientNet model is a recently developed model that uniformly scales the width, depth, and resolution of the network with a set of fixed scaling factors that are well suited for training images with high resolution. And the results of this model give a rather competitive classification efficiency, achieving 98% accuracy on the training set and 93% on the evaluation set.

## 1 Introduction

Cancer is still one of the most important problems all over the world. According to Globocan cancer statistics in 2020, the situation which people are diagnosed and died of cancer tends to increase around the world. In Vietnam, it is estimated that there are 182,563 new cases of cancer and 122,690 cancer deaths. The people who were diagnosed cancer come with a rate of 159 per 100,000 ones and the cancer mortality is 106 per 100,000 people [1].

Up to now, not only domestic researches but international ones have also frequently based on breast cancer imaging techniques to classify breast cancer which generally focuses on feature engineering before the classifier detects different kinds of breast cancer from the extracted characteristics. These features may be created manually or by the feature descriptor such as Scale Invariant Feature Transform (SIFT) [2], Generative Matrix Computation Library (GMCL) [3], Histogram of Oriented Gradients (HOG) [4], etc. For instance, Zhang and others utilized the manual feature engineering and Principal Component Analysis (PCA) to determine whether the tumor is benign or malignant [5]. Spanhol and others took advantage of machine learning methods based on the means of different feature descriptors for cancer classification [6]. Besides, Wang and others not only made use of 138 characteristic descriptors by text but they also used the Support Vector Machine to detect whether the tumor is benign or malignant [7]. Although these methods based on feature engineering gain proper accuracy for cancer classification, this process requires the wide data preprocessing, Region of Interest and hand-operated extraction which depend upon human and calculation. Furthermore, either features or the manual feature engineering is taken for a low- level feature one which may not analyze all the knowledge for concealed imaging issues such as morphology, tissue structures and other deep-tissue characteristics.

Some of the Convolutional Neutral Networks (CNN) have been demonstrated as the exceeding performance against the computer vision tasks of human. Many CNNs working paradigm are carried out for classification of biomedical images. These networks comprise AlexNet [8], ResNet [9], Inception [10], Inception-V4 and Inception-ResNet [11]. Jaffar suggested utilizing the CNN for the mammogram image classification which attained extraordinary achievement. Qiu and others [12] have operated CNN for identifying the short-term risk of breast cancer which gains the average accuracy of cancer prediction outcome as 71.40%. Rubin and Ertosun [13] have used CNN for the volume visualization observation and breast cancer classification which is up to 85% of accurate outcome. Qui and others [14] have used CNN to classify whether the tumor is benign or malignant in mammogram. Likewise, Jiao and others [15] have attained the high accurate outcome which is up to 96.7% in the same task. Sahiner and others [16] have applied CNN classifier utilizing spatial domain and texture image for mammogram classification which turned out to display 0.87 of AUC. Jadoon and others have classified 3 layers of mammogram including normal, benign and malignant which was based on the described characteristics according to CNN [17].

ConvNet models' scaling are carried out widely to get a better accuracy. For instance, ResNet may be developed from the previous versions, ResNet-18 to ResNet-200, thanks to adding more layers; recently, GPipe has attained the top 1 accuracy on ImageNet by enlarging scaling the base model 4 times up to 84.3%. However, the model scaling process for ConvNet has never been comprehensively perceived and currently there are many ways to carry out. The most popular procedure is to expand the scale of ConvNet through deep dimension or wide dimension (Zagoruyko & Komodakis, 2016). Another procedure which is less popular than the preceding one but it is becoming more and more frequent in usage is image resolution scaling [19]. Heretofore, generally, one of the three dimensions (resolution, depth, and width) was scaled. Although it is possible to scale more than one dimension arbitrarily, the arbitrary scaling is in need of an uninteresting manual scaling. Consistently, manual scaling generates sub-optimal accuracy and efficiency [18].

EfficientNets is significantly better than other ConvNets. Especially, EfficientNet-B7 has been attained the top accuracy as 84.3% which has been the most advanced model. In addition, EfficientNet-B7 is 8.4 times smaller and 6.4 times faster than GPipe. Besides, EfficientNet-B1 is 7.6 times smaller and 5.7 times faster than ResNet-152 [19].

Accordingly, we are going to carry out the research: "Histopathological Imaging Classification of Breast Tissue for Cancer Diagnosis Support" based on EfficientNet model. By using the overlapping patch method through the steps, we have effectively enhanced the image while minimizing the loss of structural and morphological features of the histopathological image. Helps EfficientNet model to be more effective because there is a combination of two factors: the model is suitable for high resolution images and trained with large data sets.

The remaining sections of our study are given as follows. In Section 2, the EfficientNet model was presented. Section 3 describes the dataset and the techniques that we used to pre-processing image. Section 4 the process of breast tissue imaging classification and building the software tool for cancer diagnosis support and Section 5 highlights the conclusion. Finally, Section 6 Acknowledgement.

## 2  EfficientNet Model

### 2.1 Model Scaling of EfficientNet

EfficientNet model, as shown in Fig. 1, has been designed on the scaling principle method of ConvNets which is possible to gain better accuracy and efficiency. Experimental study demonstrates that the most important thing is to balance all the three dimensions: width, depth and resolution. Surprisingly, the balance scaling is attained by scaling each of the three dimensions at a constant scale. Based on the observation, EfficientNet model suggests a dual scaling method which is easily manageable but efficient. Unlike other arbitrarily conventional scaling methods, the technique scales the consistent ratio among width, depth and resolution of the network with a set of permanent scaling factors. As an example, if we expect to utilize $2^N$ times as large as the computational supplies, we can easily rise the depth by $\alpha^N$ times, the width by $\beta^N$ times and the resolution by $\gamma^N$ times in which "α, β, γ" are the invariable coefficients specified by a small matrix on the initial small configuration.

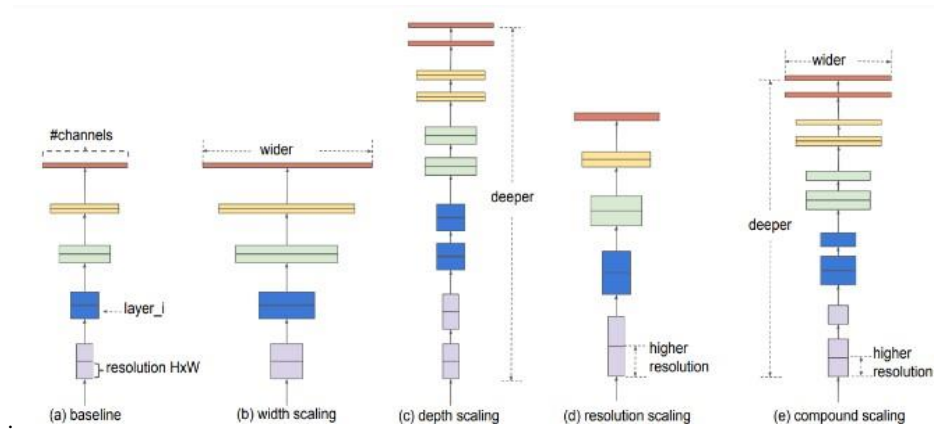

**Fig. 1.** The scaling of networks. (a) is a baseline configuration; (b)-(d) present the are conventional scales in which each model only increases one dimension according to width, depth, and resolution respectively. (e) is our proposed scaling which has uniformly scaling for all three dimensions with a fixed ratio.

## 2.2 EfficientNet Architecture

The base model of EfficientNet can be described in general in Fig.2. The processing unit of neural networks called a neuron or a node that carries out the effortless task. The node obtains the preceding input signals or an external source, then utilizes them for calculating the output signals which are propagated to the other units.

Each model contains 7 blocks, which also have a particular quantity of sub-blocks. This quantity is extended in accordance with the model versions from EfficientNet- B0 to EfficientNet-B7. The total layers of EfficientNet-B0 are 237 while EfficientNet- B7 are 813 layers. These layers are made up of the 5 modules as shown in Fig. 3.

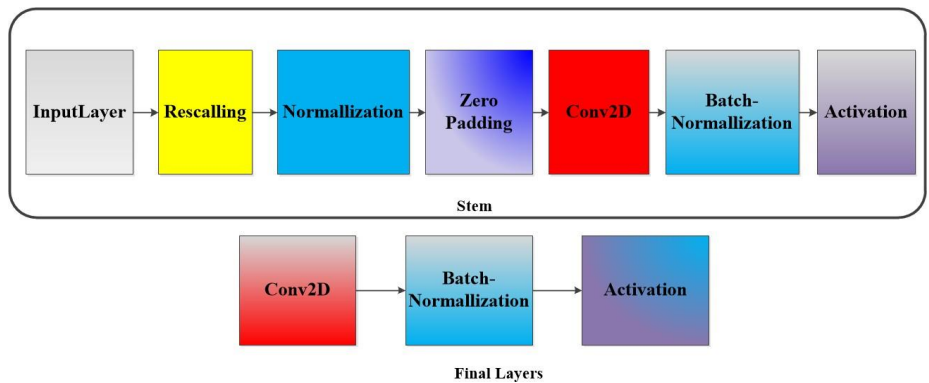

**Fig. 2.** The base model of EfficientNet.

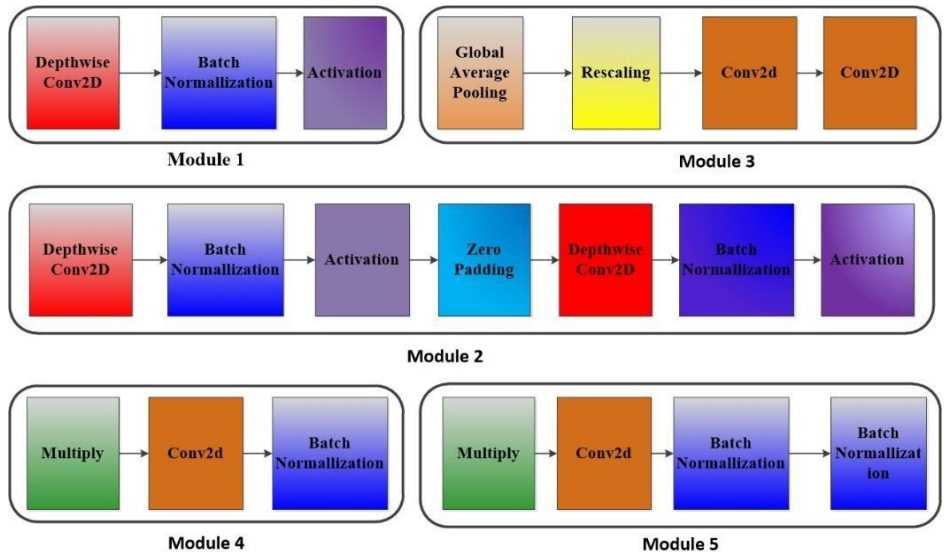

**Fig. 3.** Five modules of EfficientNet model [18].

Module 1 is made use for starting points for the sub-blocks, while module 2 is made use for starting points for the first sub-blocks of all 7 main blocks except the first one. Module 3 is utilized for connecting the blocks, while module 4 is utilized for compounding the blocks. Module 5 is utilized for compounding the sub-blocks. These modules are com- pounded for forming the sub-blocks which would be utilized for blocks in a particular way. The sub-blocks are presented in Fig. 4.

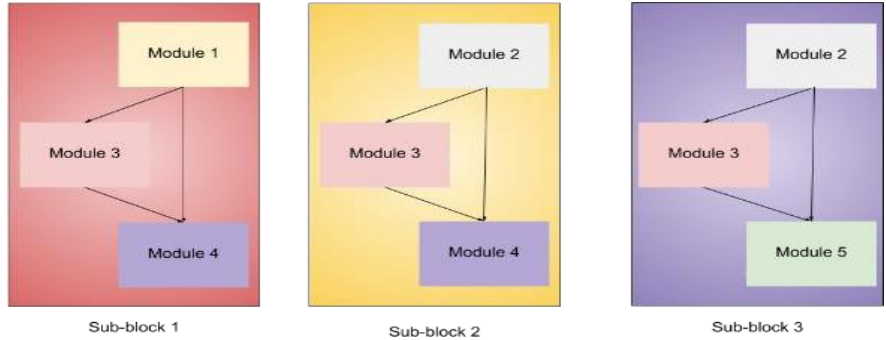

**Fig. 4.** The modules compounded for forming the sub-blocks [18].

In the first main block, the first sub-block only utilizes Sub-block 1. Sub-block 2 plays a role as the first sub-block for other main blocks. Sub-block 3 is employed for any sub-blocks except the first block in all main blocks. Increasing and compounding sub- blocks could create the EfficientNet models from B0 to B7.

## 3. Dataset and Image Pre-processing

### 3.1 Distinctive Features of Diseased Tissue Images

Breast cancer is a heterogeneous disease. This disease comprises many existence with distinct features of biology, histology and clinic. In order to comprehensively analyze malignancy in breast cancer tissues, biopsy techniques are often employed. The biopsy process (see Fig. 5.) involves collecting tissue samples, attaching them to microscope slides, then staining these slides to clearly visualize the cytoplasm and nucleus. Next, the conduct microscopic analysis of these slides was taken by pathologists to make a final diagnosis of breast cancer [20].

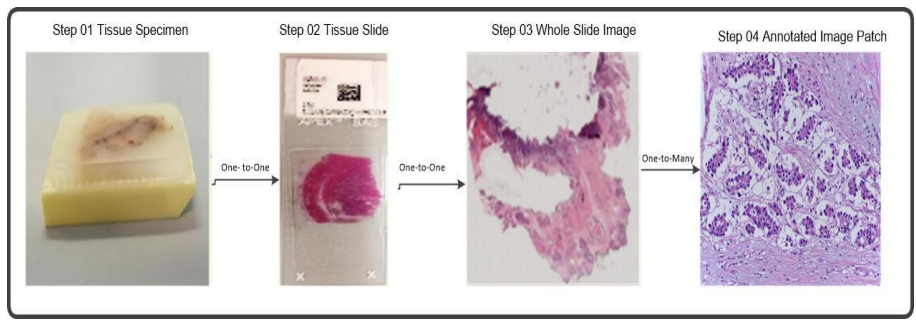

**Fig. 5.** The biopsy procedure for collecting tissue samples.

The breast tissue image is characterized by containing a lot of valuable information which is invisible in the image such as morphological information, structure information of tissues, and other characteristics of deep tissues. Specifically Kowal et al extracted 42 morphological, topological and structural features from segmented nuclei, similarly, Filipczuk et al extracted 25 shape and structure-based features from nuclei [20].

**3.2 Dataset**

The dataset of the research contains 400 images from ICIAR 2018. These images are composed of Hematoxylin and eosin (H&E) stained breast histology microscopy and its whole-slide images. These microscopic images are marked into 4 categories as depicted in Fig. 6. They are normal, benign, in situ or invasive carcinoma according to the predominant cancer in each image. This dataset contains a total of 400 microscopy images, which is distributed as given in Table 1.

**Table 1.** The features of our used dataset.

| Images with labels | Quantity | Color type | Staining |
|---|---|---|---|
| Normal (N) | 100 | RGB | H & E |
| Benign (B) | 100 | RGB | H & E |
| In Situ Carcinoma (IS) | 100 | RGB | H & E |
| Invasive Carcinoma (IV) | 100 | RGB | H & E |

We choose 70% of images for training and 20% for evaluation and the remaining 10% for test. Thus, there are 280 images for the training set, 80 images for the evaluation set and 40 images for the test set described in Table 2.

**Table 2.** Dataset division for training, validation, and test.

| | Number of Images | Data percentage |
|---|---|---|
| Training | 280 | 70% |
| Validation | 80 | 20% |
| Test | 40 | 10% |
| Total | 400 | 100% |

**3.3 Data Augmentation**

In during training process, the image data augmentation technique is used to expand a data set by generating modified images. The tensor image data was generated by Pytorch deep learning library with real-time data augmentation. The network models, which are trained with the type of data augmentation, will see new mutations of the data at each and every epochal traversal. With each input image in the batch, the model will generate a series of images through a series of translations, random rotations, etc. We set the width and height change range being 0.2 and the random rotation between [−40, 40] degrees. In the rotation operation, some pixels may be moved out of the image frame and replaced by some empty pixels, then the empty pixel will be filled through a 'reflection mode'. Finally, the batch is randomly transformed and then returned to the calling function. Table 3 presents these parameters combined with their values.

**Table 3.** Parameters of data augmentation.

| Parameters | Values |
|---|---|
| Zoom range | 0.2 |
| Rotation range | 40 |
| Width shift range | 0.2 |
| Height shift range | 0.2 |
| Horizontal flip | true |
| Vertical flip | true |
| Fill mode | Reflect |

Especially based on the method of extracting patches has proven effective in previous studies. In this paper, a sliding window of size [512x512] is applied to slide top to bottom and from left to right with 50% overlap for each step. Thus, from the original image of size [2048x1536], 35 patches were created. Such patch extraction not only collects many features of the original image, but also preserves the structure of the cells.

This data enhancement is applied to insert the image quantity for training which helps pass overfitting and describe features effectively. Especially, the patch extraction method has separated the image to be classified into 35 patches, then the image classification is based on the maximum number of types in those 35 patches.

Besides, the used dataset also contains 10 images of breast tissue collected at Cai Lay Regional General Hospital for model trial after the training. Some of these images are shown in Fig. 7.

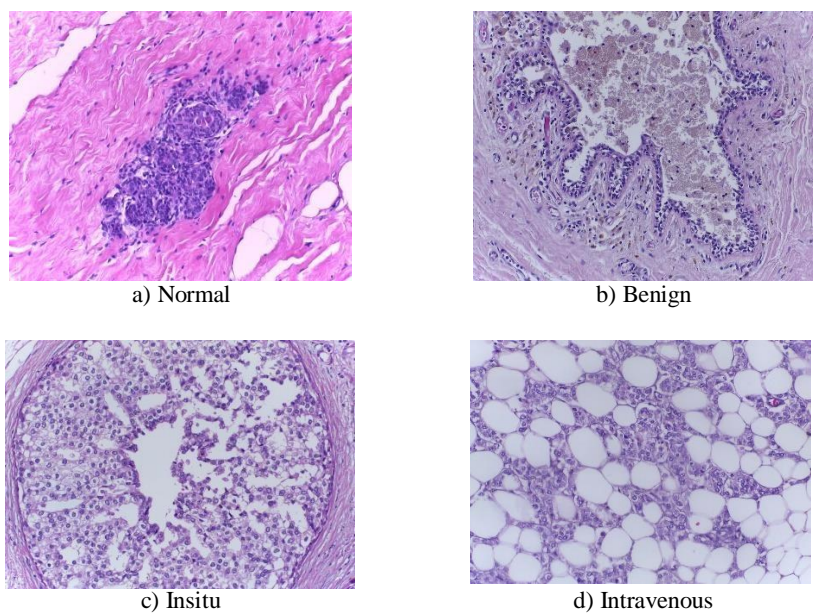

a) Normal            b) Benign

c) Insitu            d) Intravenous

**Fig. 6.** Four categories of the breast histology microscopy images

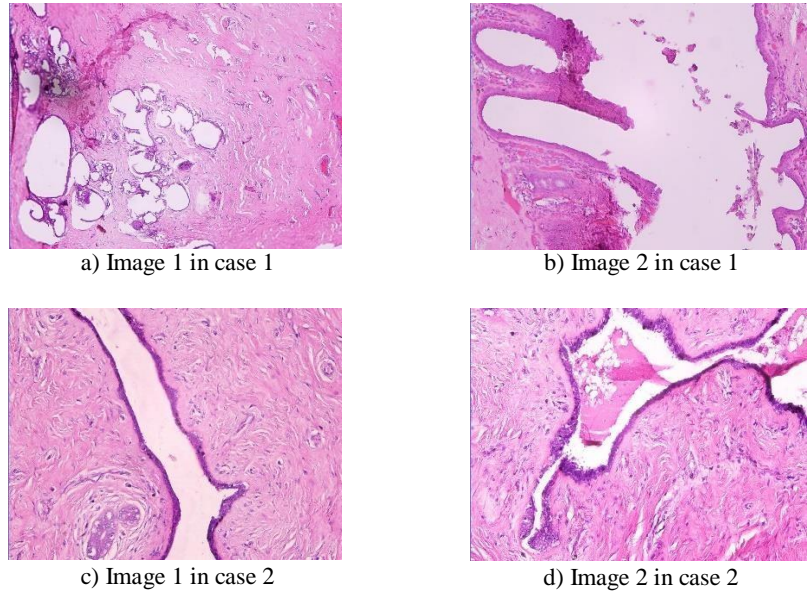

a) Image 1 in case 1    b) Image 2 in case 1

c) Image 1 in case 2    d) Image 2 in case 2

**Fig. 7.** The images collected at Cai Lay Regional General Hospital.

## 4 The Process of Breast Tissue Imaging Classification and Building the Software Tool for Cancer Diagnosis Support

### 4.1 The Process of Breast Tissue Imaging Classification

The imaging classification process is developed by using EfficientNet for the classification. Because the small data set is only suitable for small models, we only run from model B0 to model B3. Through the experimental process of EfficientNet from B0 to B3, the results have been shown that Efficient B3 is at the peak of the accuracy rate which has attained 98% on training dataset. The results are presented in the Table 4.

**Table 4.** EfficentNet models and its precision.

| Deep Learning Models | Precision |
|---|---|
| EfficientNet B0 | 86% |
| EfficientNet B1 | 90% |
| EfficientNet B2 | 91% |
| EfficientNet B3 | 98% |

We use Colab Pro to test these models. We first study the data set overview and its corresponding labels as shown in Fig.8. Then, we augment the data which is essential for the learning process. The enhancement method is detailed below in Section 4.1. Because learning requires a lot of time and limited hardware resources, we divide the learning into several stages to learn. The code format can be referred as `exp_lr_scheduler = lr_scheduler.StepLR (optimizer_ft, step_size=30, gamma=0.5)`.

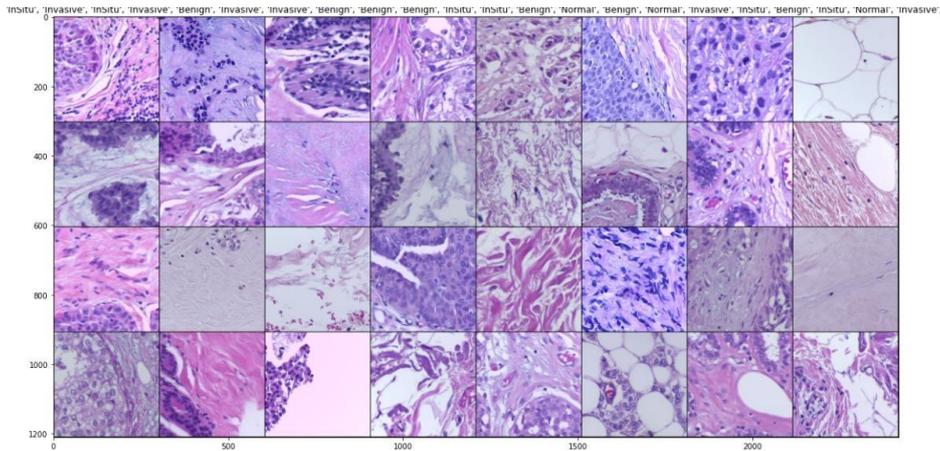

**Fig. 8.** Overview dataset and label.

**4.2 Evaluation Metrics**

The confusion matrix, also called the error matrix or contingency table, is used to evaluate the performance of our proposed model. The matrix includes four categories:

- True Positive (TP): present the number of carcinoma images which were correctly classified as carcinoma.

- False Positive (FP): present the number of non-carcinoma images which were mistakenly classified as carcinoma.

- False Negative (FN): present the number of carcinoma images which were mistakenly classified as non-carcinoma.

- True Negative (TN): present the number of non-carcinoma images which were correctly classified as non-carcinoma images.

We use precision, recall and F1-score to evaluate the classification performance of the proposed model. These measures are based on four categories of the confusion matrix in above. In that:

- Precision (P): It shows the exactness of a model which presents the ratio between the accurately classified carcinoma images to the total of predicted carcinoma images, $P = TP/{TP + FP}$

- Recall (R): It shows the completeness of a model which presents the ratio between the accurately classified carcinoma images to all carcinoma images of the dataset, $R = TP/{TP + FN}$

- F1-score (F1): It shows the harmonic average of precision and recall, $F1 = 2*P*R/{P + R}$

Table 5 and Figure 8 present measures and normalized confusion matrix of our proposed model. We use the python scikit-learn module to support in calculating these measures.

**Table 5.** Classification performance of the proposed model.

|          | Precision | Recall | F1-score |
|----------|-----------|--------|----------|
| Normal   | 0.83      | 1      | 0.91     |
| Benign   | 1.00      | 0.80   | 0.89     |
| InSitu   | 0.91      | 1      | 0.95     |
| InVasive | 1.00      | 0.90   | 0.95     |

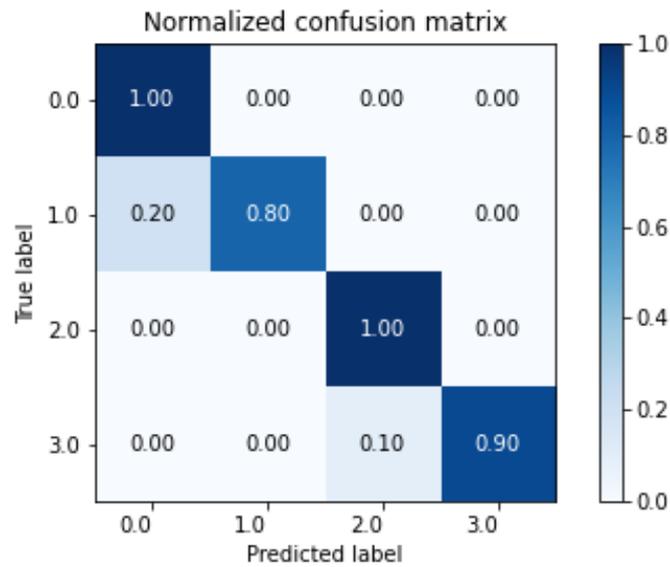

**Fig. 8.** Confusion matrix results of model B3 and measures.

### 4.3 Building The Application for Cancer Diagnosis Support

The application utilizing framework Django and EfficientNet model was trained above for histopathological imaging classification of breast tissue to support the cancer diagnosis. The application has the input and output data, and its interfaces. The input data is a histopathological image of breast tissue. The output data is the image which is included in one of the four layers: Normal (N), Benign (B), In Situ (IS), Intravenous (IV). The steps to execute web apps are as follows:

- Step 1: Enter the patient's full name, year of birth and select a photo.
- Step 2: Click View diagnostics button.

The training process was carried out on the Colab environment. After the process, we have obtained an optimal EfficientNet model in which the capability has been 4 - layer classification with the accuracy of 98%.

Creating a diagnostic support application is very effective for lower-level hospitals. It helps to answer fast results to increase patient satisfaction, in addition, it also helps doctors who are learning to practice their knowledge on the trained dataset.

## 5  Conclusion

The result shows that transfer learning backs us to generate the deep architecture of the networks with a restricted quantity of images in the training process but we have still gained the surprising result.

In addition, the enhancement of the data and the division of the original image into many small pieces stacked by 50% should also be investigated to enrich data for the training process, backing us to get over the overfitting and generating the better efficiency of the characteristics.

In the future, we will continue to enhance the application for breast cancer diagnosis support which could attain the far more accurate results. In addition, we will continue to execute all the rest of the models once more data has been collected and compare our model with other machine learning models. We will also build more features for the application to receive feedback from users and doctors to improve the model.

## 6  Acknowledgement

We thank University of Information Technology, Vietnam Aviation Academy and Ho Chi Minh City Open University for offering us a unique opportunity to accomplish the research paper.